\documentclass[11pt,a4paper]{article}
\usepackage[a4paper,top=2.5cm,bottom=2.7cm,left=2.3cm,right=2.3cm,headheight=24pt,footskip=42pt]{geometry}
\usepackage{iftex}
\ifPDFTeX
 \usepackage[T1]{fontenc}
 \usepackage[utf8]{inputenc}
 \usepackage{textcomp}
\else
 \usepackage{fontspec}
\fi
\usepackage{lmodern}
\usepackage{microtype}
\usepackage{graphicx}
\usepackage{float}
\usepackage{caption}
\usepackage{subcaption}
\usepackage{url}
\usepackage[hidelinks]{hyperref}
\usepackage{fancyhdr}
\usepackage{titlesec}
\usepackage{setspace}
\usepackage{enumitem}

\titleformat{\section}{\bfseries\large}{ }{0pt}{}
\titleformat{\subsection}{\bfseries\normalsize}{ }{0pt}{}
\titleformat{\subsubsection}{\bfseries\normalsize}{ }{0pt}{}

\begin{document}

% \begin{center}
% {\footnotesize Proceedings of the International Conference on Changing Cities VI: Spatial, Design, Landscape, Heritage \& Socio-economic Dimensions\\Rhodes Island, Greece 24--28 June 2024 $\bullet$ ISSN: \quad ISBN:}\vspace{1em}
% \end{center}
\begin{center}
{\bfseries\Large Revitalizing Public Urban Places through Cultural and Political Memory: A Technological Approach with LLMs and Augmented Reality}\par
\vspace{0.8em}
{\bfseries L. Vartziotis\textsuperscript{1,2*}, T. Vartziotis\textsuperscript{1,2}, V. Keckeisen\textsuperscript{2}, F. Beutenmueller\textsuperscript{2}, M. Obstbaum\textsuperscript{2}, S. Kotsopoulos\textsuperscript{1,3}, K. Moraitis\textsuperscript{1}}\par
\textsuperscript{1}National Technical University of Athens, Greece\par
\textsuperscript{2} TWT GmbH Science \& Innovation\par
\textsuperscript{3} Massachusetts Institute of Technology \par
\vspace{0.3em}\emph{*Corresponding author: E-mail: lara.vartziotis@twt-gmbh}\par
\end{center}

\section*{Abstract}
This paper delves into the intricate intersection of memory, place, and identity, exploring how new technologies, particularly Apple Vision Pro, illuminate this nexus. Leveraging the capabilities of digital twins and virtual reality, we embark on a multifaceted exploration to deepen our comprehension of how memory is intricately woven into landscapes and urban environments of cultural and historical significance. Through this approach, we discern key visual elements that evoke memory and heritage within physical environments.

Innovative applications through tools such as Apple Vision Pro can facilitate image extension to define place identity, where the technology informs the viewer on cultural and political entities of a varying timeline. Concurrently, visual storytelling techniques curated with Apple Vision Pro showcase the dynamic evolution of landscapes and the preservation of cultural heritage, offering new insights into the interplay between memory and physical environments. Furthermore, the integration of Virtual Reality (VR) technologies enables the recreation of historical landscapes and urban-scapes. This immersive approach invites users to transcend temporal boundaries, immersing themselves in reconstructed environments and experiencing the past in a dynamic and engaging manner. Semantic Image Search capabilities implemented using Apple Vision Pro streamline research efforts, facilitating the exploration of mnemonic landscapes. By uncovering images related to specific themes or keywords such as monuments, tradition, or cultural identity, we deepen our understanding of the intricate connections between memory, place, and identity.

This research investigates a methodology to connect digital twins and virtual environments with urban and non-urban environments and landscapes to illustrate the concepts of cultural, historical, and environmental sustainability. Central to this approach is defining the resilience of the current state, its evolution into the future, and emphasizing the significance of the past. These technologies not only facilitate a historical and cultural embrace but also evoke a nostalgic feeling of returning to a specific place years later. To analyze and quantify this approach to identity and its evaluation over time, a methodology is introduced. This methodology outlines the integration of technologies necessary to achieve the goal of revitalizing public urban places through cultural and political memory.

Through these innovative applications, this paper contributes to the advancement of research in digital twins of spaces, urban transformation, and cultural heritage preservation. By offering novel insights into the complex relationship between memory, place, and identity in the digital age, we pave the way for a deeper understanding of our collective past and its impact on the present.

\noindent\textbf{Keywords:} AR; identities; NTUA; TWT, Apple Vision Pro; Visual Reality; Immersive Experiences; Augmented Reality

\section*{1. Introduction}
Virtual and augmented reality environments present nowadays an interest in researchers, engineers, and architects because of the established visual production and communication of ideas in modern culture. The same attention is given to the newly introduced technology of large language models (LLMs) with the commonly accepted tool ChatGPT. This paper briefly shows one of the latest developments in the field of artificial intelligence and its potential applications to the fields of architecture and space representation. This study briefly discusses how LLMs can be effectively used to understand the space in which we are located, whether it is a square, a monument, a landscape, or a building of significant importance. LLMs are artificial intelligence systems that generate text or translate visual images using a generative pre-trained transformer (GPT) architecture [1]. The paper presents some possible use cases for LLMs in urban understanding. Given the benefits of LLMs, they are also prone to limitations such as biased results, prompt injection attacks, and the colossal requirements of computational resources [1]. It is to be noted that the LLMs are helpful but still in the early stages of development and need further in-depth research and evaluation for efficient use in transportation systems [1]. This work aims to address the integration of such technologies into the infrastructure and, more precisely, the identity of a place. Recent advancements in multimodal large language models (MLLMs) have achieved significant multimodal generation capabilities, akin to GPT-4, [2]. These models predominantly map visual information into language representation space, leveraging the vast knowledge and powerful text generation abilities of LLMs to produce multi-modal instruction-following responses [2]. Specifically, we introduce the augmented reality connected to the multimodal generation capabilities of LLMs to understand, illustrate, and predict the identity of a place.

Augmented reality (AR) merges computer-generated data with the real world, enhancing people's perception of reality by integrating synthetic information into the live view of the physical environment. In AR, the physical world retains its predominant role, with computer-generated elements complementing and enriching the user experience. This technology falls within the realm of mixed reality [16], where virtual and real-world objects coexist seamlessly in the same space. AR systems such as the Apple Vision Pro enable real-time interaction by combining real and virtual elements within a real environment, while also accurately registering virtual objects in three-dimensional space.

Genuine environments, whether urban or periurban landscapes, are undeniably fundamental to our existence. However, they are not isolated entities merely existing in tangible form. Reality and its material expression are always accompanied, sustained, and perpetuated by intangible mental phenomena such as memories, fantasies, desires, and even fears, which manifest in symbolic or semantic expressions [15]. Over the past decade, AR has presented intriguing opportunities and practical applications for revealing cultural heritage, promoting historical materials, and facilitating interactive visualizations of heritage items [17].

Archeoguide was one of the first AR systems that was built at the archaeological site of Olympia in Greece from 2000–2002 [18]. It provided personal AR tours and reconstruction of the ruined cultural heritage sites to help people understand and experience the past. The equipment that has been used to make the 3D of the monuments possible is a head-mounted display (HMD) with an external camera and a compass, a backpack with a computer, a battery, and wireless communication equipment [18].

Highlighting the significance of public urban and peri-urban spaces as catalysts for social identity and cohesion, the municipality of Kaisariani, situated on the outskirts of Athens, presents another compelling project. The redesign of the square aimed not only to provide visitors with recreational opportunities but also to offer them the chance to explore a virtual narrative of the historical and cultural heritage of the surrounding area. This initiative intertwines physical and virtual networks, offering visitors within the physical space and beyond the opportunity to engage with the cultural landscape through primarily virtual navigation. The creation of this virtual visiting network enables continuous data collection, enriching both the virtual tour structure and future utilization of the physical space. The square's surface was transformed into a map depicting the eastern part of the Aegean Sea and the western coastal zone of Minor Asia [15]. Metallic signs embedded in the square floor feature QR codes, inviting visitors to embark on a virtual journey through the history of the Hellenic communities in Minor Asia [15]. Through such initiatives, which emphasize social interactions and the public display of collective affiliations, the crucial role of public spaces in shaping identities becomes evident.

\section*{2. Identity of Place Definition}
Place identity, as a fundamental concept in human-environment relations, encompasses a variety of psychological theories and dimensions. Initially, place identity is defined as the aspects of self that shape an individual's personal identity in relation to the physical environment through a complex interplay of conscious and unconscious thoughts, emotions, values, and behavioral inclinations [5]. Various studies have further explored this concept, establishing connections between it, subjective regional identity, and attachment to places. Place identity has been described as a combination of physical processes, man-made elements, and the meanings attributed to places, emphasizing the diverse components that contribute to a place's identification within the spatial system [6]. They perceive the identities of places in varied ways, distinguishing them by considering diverse elements like physical features, cultural attributes, historical associations, experiential ties, and more [7]. There are mediating variables that impact place identity during the transition from the past, present, and future.

Identity consists of two primary components. The first is the intensity, which signifies the strength of belief, and the second is the degree of positive connection with the object of identity [13]. Overall, place identity serves to distinguish regions from one another, drawing on elements such as nature, culture, inhabitants, language, and historical connections [7].

\subsection*{2.1 Evolution of Place Identity}
The evolving nature of society, characterized by technological advancements, urbanization, and globalization, continually reshapes the physical and cultural landscapes of places, altering the ways individuals perceive and engage with their environments. Amidst these transitions, individuals navigate the complexities of past experiences, present realities, and future aspirations, shaping their evolving sense of place identity. Place identity, conceived as a subjective social construct, emerges as individuals distinguish one place from another, drawing on subjective cognition and objective physical settings. Memory plays a pivotal role in this process, serving as a mechanism that intertwines sensation with intellect and links present conditions with past experiences [10]. As memory intertwines with place, it becomes evident that places acquire identity not solely through material substance but also through intangible processes, collective memories, and ideological formations [10]. This dynamic interplay between memory and place underscores the significance of the "emplacement" of memory, highlighting the intricate relationship between collective memory and the formation of place identity [10].

\subsection*{2.2 Transforming space identity in society and architecture.}
The dynamic nature of societal changes and architectural evolution has significant impacts on both society and the architectural community. The swift progression of urbanization and globalization in society necessitates a reevaluation of cultural identity and social interconnectedness. As communities embrace new influences and experience architectural changes, old values may undergo a transition, which can affect social unity and collective memory. Furthermore, political interventions and global events such as globalization have the potential to disturb the historical flow and redefine the narrative of a location's identity. These changes force communities to balance the conservation of their cultural traditions with the need for progress, impacting the citizens' sense of identity and ongoing connection.

These shifts require the architectural community to reassess design ideas and approaches to creating places. Architects are faced with the challenge of balancing the preservation of historical identities with the accommodation of modern demands and aspirations. This necessitates a meticulous equilibrium between conservation efforts and inventive design interventions that protect cultural heritage while promoting adjustment to contemporary circumstances. Moreover, the impact of worldwide trends and international cooperation significantly influences architectural discussions, resulting in the development of novel paradigms and design methodologies.

Environmental psychology is crucial for comprehending human behavior in constructed environments [12], assisting architects in designing spaces that evoke people's emotions, values, and memories. The analysis and improvements of current theories in environmental psychology provide guidance for design techniques that prioritize well-being and user experience. This reflects the shifting dynamics of place identification in a rapidly evolving world [11]. Over time, the alteration of a community's surrounding natural environment reshapes their social structure, consequently impacting the place's identity [11].

\section*{3. Technological Tools: AR Technology \& LLM integration}
\subsection*{3.1 AR and VR technologies}
In this section, the evolution, applications, and emerging challenges of augmented reality (AR) and virtual reality (VR) are outlined, identifying their transformative impact on place and identity.

AR technology enriches the real world by overlaying digital content like images and sounds onto the user's environment in real time, seamlessly integrating digital elements into their surroundings. Conversely, VR immerses users in simulated environments, disconnecting them from the physical world through devices like VR headsets that control visual and auditory perceptions. This immersive experience enables users to explore various environments and scenarios, ranging from lifelike to fantastical realms [23].

While definitions of AR and VR were more clearly defined and popularized in the 1990s, studies about these technologies were already found in the late 1950s [23]. Early AR and VR systems struggled with accessibility and practical use. For example, the ARCHEOGUIDE project, an early 2000s initiative funded by the EU to enhance archaeological site tours with AR, involved the use of head-mounted displays, a laptop with several hardware attachments, as well as heavy backpacks, which highlights the significant logistical complications of early AR implementations [26].

In recent years, AR and VR have found widespread application, for example, in architectural design and education [19]. Today, these technologies are increasingly integrated into infrastructure projects to enhance visualization [22] of complex infrastructure systems [23] or for guided tours in museums [24] [20].

Despite advancements in AR and VR, challenges remain in the widespread adoption of AR and VR in said fields. Issues include complications in outdoor environments due to temperature and luminance changes, device user friendliness [21], localization method accuracy [21], practicality, and high latency [23].

The latest technology, like the Apple Vision Pro, a mixed reality headset, offers high-resolution displays and spatial audio capabilities, enhancing the user experience. The Apple Vision Pro incorporates eye and hand tracking technologies by using a total of twelve cameras, six microphones, and five sensors to enable intuitive interaction with virtual environments without the need for conventional controllers. Two onboard processors, one specifically designed to calculate inputs, are included to reduce latency, making it possible to transmit sounds and images within 12 milliseconds, which is an important aspect of reducing motion sickness [25].

Future research in AR technology may focus on addressing challenges like occlusion, localization method accuracy [22], enhancing device affordability and performance, and developing standardized AR tools for better application integration and comparison across academic and engineering communities [24].

\subsection*{3.2 Augmented reality in the infrastructure world}
Apple Vision Pro overlays digital content onto real-world environments. While standing at Monastiraki Square in Athens, one can view the surrounding area and simultaneously access historical images and information from outputs like ChatGPT or other large language models (LLMs). Apple Vision Pro overlays digital content onto real-world environments. While standing at Monastiraki Square in Athens, one can view the surrounding area and simultaneously access historical images and information from outputs like ChatGPT or other large language models (LLMs).

The device’s spatial mapping ensures that digital elements are accurately placed onto real-world objects and stay fixed in the user's physical surroundings, as it has the ability to adapt to the physical environment by using, for instance, a LiDAR (Light Detection and Ranging) sensor. Furthermore, Vision Pro makes it easy to interact with multiple windows, as the user can open various applications and information sources and is able to place them around himself. Precise eye tracking enables intuitive navigation, as one can select virtual content by simply looking at it. Hand gesture recognition is another key feature, allowing users to, for example, resize, move, or close application windows with simple gestures. Besides, voice control and Siri integration provide hands-free operation by allowing users to issue commands and search for information [25].

All these features make it easier and more accessible for users to connect real-time information from LLMs like ChatGPT with locations, creating a deeper understanding of place and identity.

The device’s spatial mapping ensures that digital elements are accurately placed onto real-world objects and stay fixed in the user's physical surroundings, as it has the ability to adapt to the physical environment by using, for instance, a LiDAR (Light Detection and Ranging) sensor. Furthermore, Vision Pro makes it easy to interact with multiple windows, as the user can open various applications and information sources and is able to place them around himself. Precise eye tracking enables intuitive navigation, as one can select virtual content by simply looking at it. Hand gesture recognition is another key feature, allowing users to, for example, resize, move, or close application windows with simple gestures. Besides, voice control and Siri integration provide hands-free operation by allowing users to issue commands and search for information [25].

All these features make it easier and more accessible for users to connect real-time information from LLMs like Chatgpt with locations, creating a deeper understanding of place and identity.

\subsection*{3.3 LLMs and Generative AI in the infrastructure world}
Large language models (LLMs) are AI systems trained on massive datasets that enable them to perform complex language tasks. With billions of parameters, LLMs can infer context, generate responses, translate languages, summarize text, answer questions, and even assist with creative writing and code. These capabilities are revolutionizing fields like chatbots, virtual assistants, content creation, research, and translation, making LLMs a cornerstone of modern AI. Large language models (LLMs) are a category of foundation models trained on immense amounts of data, making them capable of understanding and generating natural language and other types of content to perform a wide range of tasks [3]. LLMs have become household names thanks to the role they have played in bringing generative AI to the forefront of the public interest [3]. This technological advancement has occurred alongside the evolution of machine learning, machine learning models, algorithms, neural networks, and the transformer models that provide the architecture for these AI systems. LLMs are a class of foundation models that are trained on enormous amounts of data to provide the foundational capabilities needed to drive multiple use cases and applications [3].

LLMs represent a significant breakthrough in natural language processing (NLP) and artificial intelligence and are easily accessible to the public through interfaces like Open AI’s Chat GPT-3 and GPT-4, Meta’s Llama models and Google’s Gemini tool, IBM’s Granite model with the products Watsonx Assistant and Watsonx Orchestrate.

\subsection*{3.3 LLM Tools to describe place identity: An approach with ChatGPT.}
ChatGPT, when combined with DALLE, showcases numerous notable abilities in generating outcomes. DALLE, the AI system that creates realistic images and art from a description in natural language, is directly integrated into ChatGPT [4]. ChatGPT's advanced natural language understanding is demonstrated by its capacity to comprehend and interpret intricate prompts regarding historical eras and hypothetical futures. The generation of intricate prompts for DALLE exemplifies ChatGPT's aptitude for imaginative cognition of the future development of the space. It creates narratives by combining historical facts with imaginative elements that are appropriate for a specific time period. The integration of ChatGPT with DALL·E showcases the cross-modal capacity of AI systems to convert textual descriptions into visual representations. ChatGPT demonstrates its profound understanding of historical facts by integrating well-known historical aspects of a specific space into the image generation prompts, showcasing its comprehensive knowledge of the past up to the present time. The tool has the ability to demonstrate urban developments based on the AI's proficiency in generating realistic future scenarios. The diverse functionality of AI is demonstrated through its ability to be utilized for educational, creative, and speculative purposes. Additionally, we emphasize AI's ability to connect the divide between written and visual communication, thereby creating fresh opportunities for generating content and ideas.

\section*{4. Methodology and Discussion}
\subsection*{4.1 Methodology}
This section outlines our approach for evaluating the identification of spaces. Initially, we determine the specific location or space that we wish to analyze, either by capturing our own photograph of the space or by locating it through Google Street View. In this section, we provide a detailed explanation of the methodology employed to establish the identity of the LLM tool. Finally, we present the incorporation of this data into the Apple Vision Pro to enhance the current environment with the identification produced by ChatGPT. Figure 1 depicts the concept of place identity and how it is visually represented through the use of technology.

Based on the work performed by ChatGPT and DALLE in creating and processing complicated prompts and visualizing information, we created a methodology pipeline. To carry out our preliminary assessment, we needed a collection of input data from which we could derive context. During the first stage of our method, we decided on using rapid formulation. During the first stage of our method, we opted to utilize prompt formulation.

Concretely, we obtained the image of Google StreetView and instructed ChatGPT to ascertain whether it recognizes the space. We provide the name of the space as additional context and information in situations where recognition fails. Subsequently, we depended on a crucial element that involved creating clear and accurate questions to obtain information. This involved skillfully employing natural language processing (NLP) techniques to analyze and organize user inputs into clear prompts, address their informational requirements related to space identification, and provide historical and cultural insights.

As mentioned before, ChatGPT is trained using an extensive amount of data, which encompasses photographs and historical records. LLMs are employed to measure identity by soliciting space-related information, such as inquiries about the construction date of a square, the identity of its builder, the evolution of its structure and its usage from ancient to Byzantine to modern times, and the architectural trends that influenced the emergence of new standards and design approaches during that period.

The utilization of DALLE is a result of combining Apple Vision Pro with the visual depiction of space. An interaction involving the utilization of Apple Vision Pro in Google StreetView, along with a real-time conversation with ChatGPT, whether through written or spoken language, simulates a discussion with someone who possesses knowledge about the subject matter, like a tour guide equipped with comprehensive information or gained expertise. Taking this a step further, we aim to convert the textual information into a visual representation, specifically an image. We request DALLE to develop an image depicting the spatial changes over various time periods throughout the years. We specifically requested an image depicting Monastiraki Square during three distinct periods: ancient times, the Byzantine era, and a speculative future scenario set 50 years after the completion of this project, in 2074.

We proceed with an iterative process of improvement, recognizing that the provided images are not derived from accurate historical images and sources but rather serve as a means of generating imaginative representations. Therefore, we proceed with the feedback loop, which enables the user to observe the produced image and offer input regarding its precision and aesthetic attributes. The main issue with these technologies lies in their limited accuracy in generating photographs and their inability to produce photos from historical sources, instead relying on generating them from scratch. However, it should be noted that ChatGPT offers references to websites where users can access and review information about the space. The final stage of the process focuses on delivering the output and facilitating user interaction. We accomplish this by presenting a polished image to the user through a user-friendly interface, specifically the Apple Vision Pro. Afterwards, users are able to engage in an interactive exploration using the Apple Vision Pro, enabling them to interact with the generated content. This includes actions such as zooming, editing, or overlaying supplementary information, which enhance the educational and creative experience. Figure 2 illustrates the methodology of place identity, which is based on a technological approach involving LLMs and augmented reality.

This pipeline emphasizes the integration of NLP and computer vision, enabled by machine learning methods, to not only produce spatial identification of content but also create an interactive and dynamic platform for creative and instructional objectives. The pipeline is designed to be iterative, with user feedback immediately influencing continuing enhancements and adjustments, ensuring that the system adapts to user needs and technical changes.

\begin{figure}[H]
\centering
\includegraphics[width=0.9\linewidth]{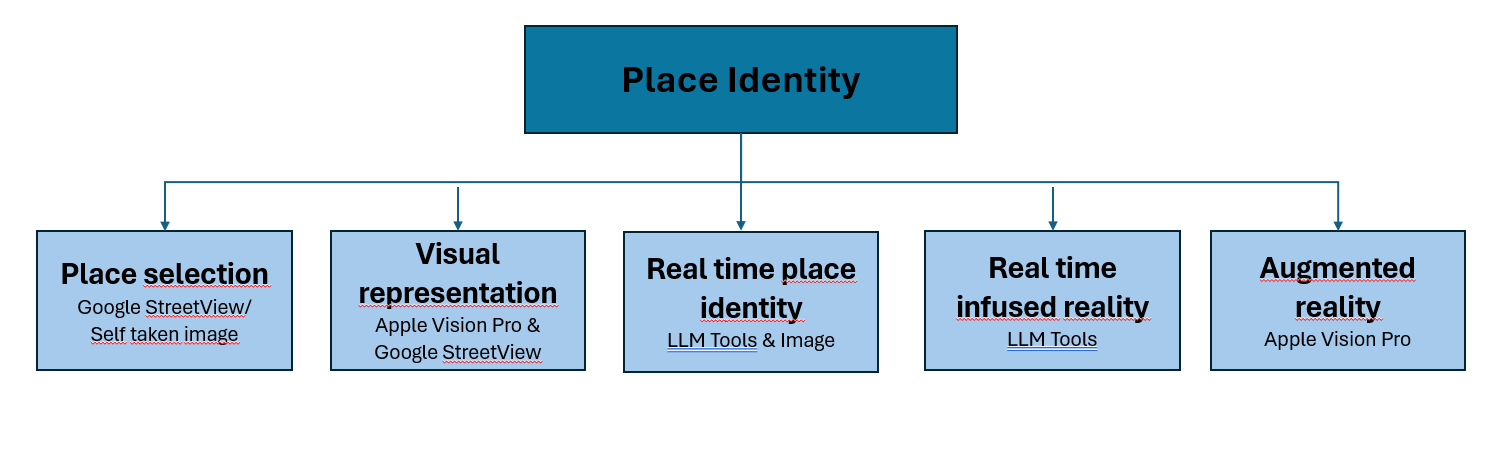}
\caption{Place identity and its representation by technology}
\end{figure}

\begin{figure}[H]
\centering
\includegraphics[width=0.9\linewidth]{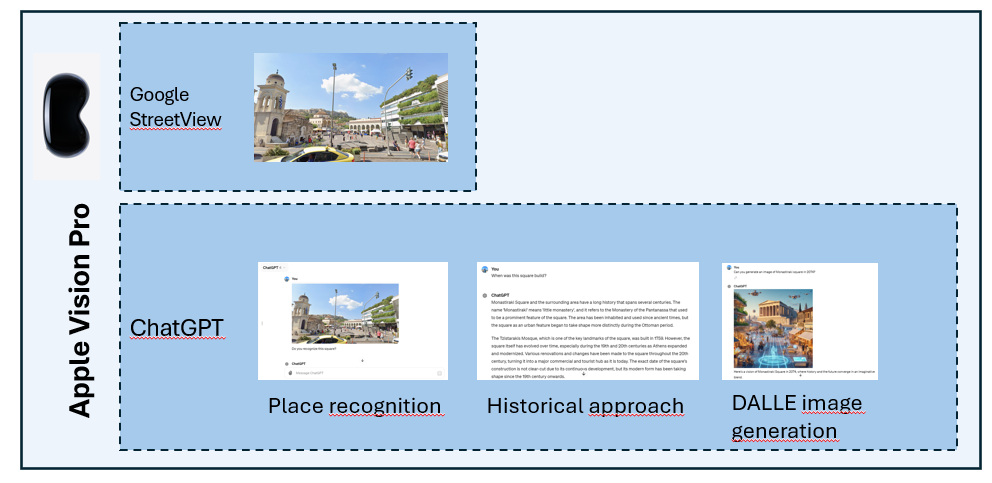}
\caption{The methodology of place identity based on the technological approach with LLMs and augmented reality}
\end{figure}

\subsection*{4.2 Use cases and Discussion}
For this methodology, we explored the Monastiraki Square, which is one of the most vibrant squares in Athens, Greece, with a long history of architectural, historical, and cultural importance. ChatGPT immediately recognized the square and was able to answer all the questions we asked. The wanted approach would have been to discuss with ChatGPT as if it were a real tour guide or a historian and architect all together. It gives us information on “the mixture of architectural styles, including the Tzistarakis Mosque with its distinctive dome, which now houses part of the Museum of Greek Folk Art, and the Byzantine Church of the Pantanassa." We also received cultural information: “Monastiraki is known for its flea market and vibrant atmosphere, a mix of ancient and modern that’s characteristic of Athens." When asked, “Can you show me pictures from the ancient time of the square?” ChatGPT answered: “Finding authentic pictures of Monastiraki Square from ancient times would be quite challenging, as the concept of photography did not exist until the 19th century. However, there are historical records and drawings that provide us with an understanding of what the area may have looked like.” and was not able to show any illustrations of the square. “I'm sorry for the confusion, but as an AI developed by OpenAI, I currently don't have the ability to browse the internet or access external databases in real-time, so I can't retrieve or show existing images or drawings from external sources.”

To build a visual understanding of these times, we requested the representation of these eras as shown in Figure 2. The goal is to correlate the different time periods with the space identification and its evolution. More accurate results could be obtained by combining tools such as ChatGPT and Gemini. While Gemini does not recognize the square, with its current technology, it is able to show a search in the network and accurately find an ancient representation of Athens, as given by https://ancientathens3d.com/, a similar approach someone would follow on a Google search. The result of this approach is shown in Figure 3.

\begin{figure}[H]
\centering
\begin{subfigure}{0.31\linewidth}
\centering
\includegraphics[width=\linewidth]{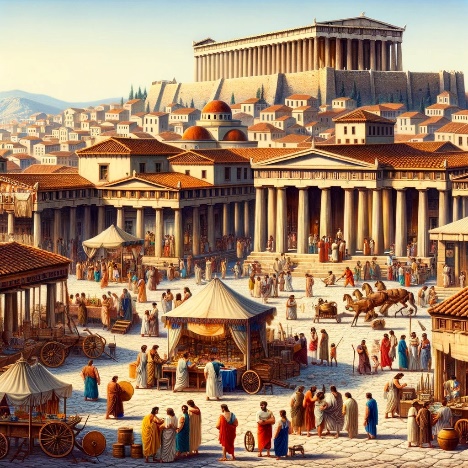}
\caption*{(a) Ancient}
\end{subfigure}\hfill
\begin{subfigure}{0.31\linewidth}
\centering
\includegraphics[width=\linewidth]{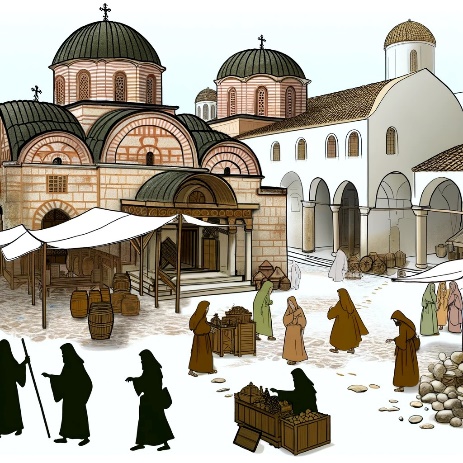}
\caption*{(b) Byzantine}
\end{subfigure}\hfill
\begin{subfigure}{0.31\linewidth}
\centering
\includegraphics[width=\linewidth]{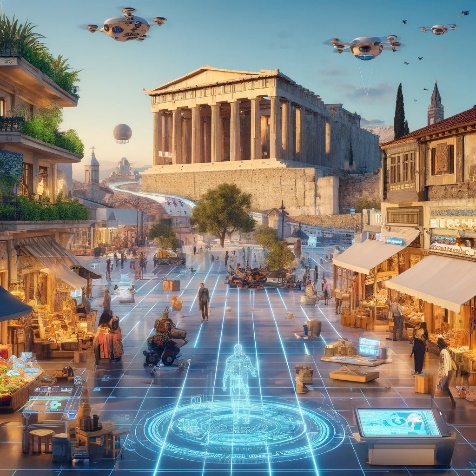}
\caption*{(c) Future}
\end{subfigure}\hfill
\caption{DALLE generated images on three different representations of the Monastiratki square.}
\end{figure}

\begin{figure}[H]
\centering
\includegraphics[width=0.65\linewidth]{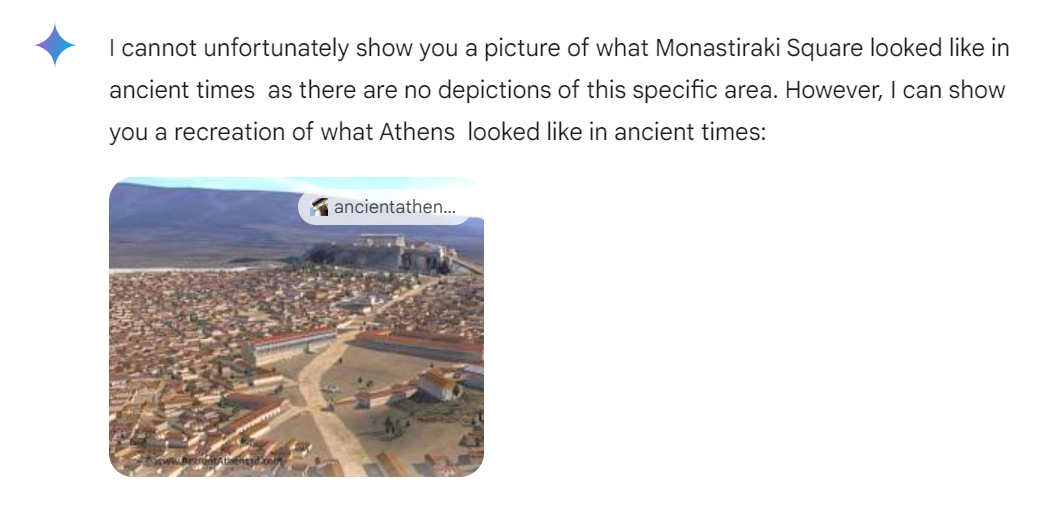}
\caption{Gemini provided representation of Monastiraki square}
\end{figure}

% Figure 5. The space visual represented with Apple Vision Pro

In these examples we show the powerful ChatGPT tool but as shown and described in the methodology there are a few advancements that need to be considered. For instance, there is a need for customized training models for these tasks. During the work of this pipeline, we show the importance of the improvement of the LLM tools to adjust their capabilities on space identity, and Llama is the most capable openly available LLM [4]. The significance of the Llama tool by Meta is the fact that it allows further training on new more specific tasks. The vision is to enable developers to customize Llama 3 to support identification analysis and historical correlation of places. More specifically the LLM model can be re-trained, specializing in the space's identity and its historical, cultural, and political transformation throughout time.

\section*{4. Conclusion}
This study has critically explored the transformative potential of augmented reality (AR) and large language models (LLMs) in redefining public urban spaces through the lens of cultural and political memory. By integrating AR with the sophisticated capabilities of LLMs, particularly through platforms like Apple Vision Pro and ChatGPT, we have outlined a robust methodology to enhance the identification and appreciation of place identity in urban environments. Our findings demonstrate that AR can significantly enrich the physical experience of a location by merging digital information with real-world contexts, thus allowing for a deeper engagement with the historical and cultural layers of urban spaces. Virtual Reality (VR) further extends these capabilities by creating immersive environments where past, present, and future narratives of urban landscapes can be experienced, offering users a unique perspective that is not possible through traditional means. The use of LLMs, exemplified by ChatGPT, has shown promising results in generating detailed narratives and visual representations that anchor the memory of a place within its physical attributes. These models not only facilitate a richer interaction with the environment but also act as repositories of cultural memory, capable of conveying complex historical and social narratives through their generative abilities. The integration of these technologies into the study of place identity suggests a dynamic and innovative approach to urban design and heritage conservation. The ability to overlay digital twins with real environments in AR, combined with the narrative and analytical power of LLMs, provides a powerful tool for architects, planners, and heritage professionals to explore and interpret urban spaces in ways that were previously unattainable. Future research should focus on refining these technological tools, improving their accuracy, user-friendliness, and accessibility. Additionally, further exploration into the ethical implications and potential biases inherent in AI-driven interpretations of cultural and political memories is crucial. In conclusion, this research paves the way for a new era of urban and architectural design, where digital and physical realms merge to enhance our understanding of place identity. By harnessing the power of AR and LLM technologies, we can not only preserve but revitalize the memory embedded within urban landscapes, ensuring that they remain dynamic, relevant, and meaningful for future generations.

\section*{References}
\begin{enumerate}[leftmargin=*,label={\arabic*.}]
\item Sharma, Ankit \& Vashist, Komal \& Aggarwal, Praveen \& Sachdeva, Som Nath. (2023). Use of LLMs in Transportation Engineering: A Brief Review.
\item Li, Y., Hu, B., Wang, W., Cao, X., \& Zhang, M. (2023).
Towards Vision Enhancing LLMs: Empowering Multimodal Knowledge Storage and Sharing in LLMs.
\textit{arXiv preprint arXiv:2311.15759v1}.
\url{https://arxiv.org/html/2311.15759v1}
\item IBM. (n.d.). What Are Large Language Models (LLMs)? Retrieved [01.05.2024], from \url{https://www.ibm.com/topics/large-language-models}
\item \url{https://ai.meta.com/blog/meta-llama-3/}
\item Proshansky H. M., Fabian A. K. (1987). “The development of place identity in the child,” in The Built Environment and Child Development, eds Weinstein C. S. David T. G. (New York, NY: Plenum Press; ), 21–40.
\item Groote P., Haartsen T. (2008). “The communication of heritage: creating place identities” in The Ashgate Research Companion to Heritage and Identity, eds Graham B., Howard P. (Hampshire: Ashgate Publishing; ), 181–194.
\item Peterson G. (1988). Local symbols and place identity: tucson and albuquerque. Soc. Sci. J. 25 451–461. 10.1016/0362-3319(88)90024-9
\item Paasi A. (2002c). “Regional identities and the challenge of the mobile world,” in Kulturell Identitet og Regional Utvikling, ed. Engen T. O. (Elverum: Høgskolen i Hedmark;), 33–48.
\item Siemens, Reynold L. (1988). Hegel and the Law of Identity. The Review of Metaphysics. 42 (1): 103–127. ISSN 0034-6632. JSTOR 20128696.
\item Verykokou, Styliani \& Ioannidis, Charalabos \& Kontogianni, Georgia. (2014). 3D Visualization via Augmented Reality: The Case of the Middle Stoa in the Ancient Agora of Athens. 10.1007/978-3-319-13695-0\_27.
\item Al-Mohammedy, Leena \& Al-Nashmi, Njoud \& Baabdullah, Renad \& El-Shorbagy, Abdel-Moniem \& Taylor, George. (2022). Emergence of New Place Identities through Architecture. Civil Engineering and Architecture. 10. 1590-1598. 10.13189/cea.2022.100428.
\item K. Franck. Exorcising the ghost of physical determinism. Environment and Behavior, Vol. 16, No. 4, 411-435, 1984.
\item Barney, J.B.; Bunderson, S.; Foreman, P.; Gustafson, L.T.; Huff, A.S.; Martins, L.L.; Stimpert, J.L. A strategy conversation on the topic of organization identity. In Identity in Organizations. Building Theory Through Conversations; Whetten, A.D., Godfrey, P.C., Eds.; Sage Publications: New York, NY, USA, 1998; pp. 99–168. [Google Scholar]
\item "DALL·E Now Available in Beta". OpenAI. 20 July 2022. Archived from the original on 20 July 2022. Retrieved 20 July 2022.
\item Chelidoni, A., \& Moraitis, K. (2022). Smart cultural and political narratives in urban and periurban landscape. Technical Annals, 1(1), 271–280.
\item Milgram, P., Takemura, H., Utsumi, A., Kishino, F.: Augmented Reality: A class of displays on the reality - virtuality continuum. In: Telemanipulator and Telepresence Technologies Conference of the SPIE International Symposium on Photonics for Industrial Applications, pp. 282—292. SPIE, Washington (1994)
\item Foni, A. E., Papagiannakis, G. Magnenat-Thalmann, N.: A Taxonomy of Visualization Strategies for Cultural Heritage Applications. ACM Journal on Computing and Cultural Heritage 3, 1—21 (2010)
\item Vlahakis, Vassilios \& Karigiannis, John \& Tsotros, Manolis \& Gounaris, Michael \& Almeida, Luís \& Stricker, Didier \& Gleue, Tim \& Christou, Ioannis \& Ioannidis, Nikolaos. (2001). ARCHEOGUIDE: first results of an augmented reality, mobile computing system in cultural heritage sites. 131-140. 10.1145/584993.585015.
\item Cipresso, P., Giglioli, I. A. C., Raya, M. A., \& Riva, G. (2018). The past, present, and future of virtual and augmented reality research: A network and cluster analysis of the literature. Frontiers in Psychology, 9, 2086.
\item Minaee, S., Liang, X., \& Yan, S. (2022). Modern augmented reality: Applications, trends, and future directions. arXiv preprint arXiv:2202.09450.
\item Tan, Y., Xu, W., Li, S., \& Chen, K. (2022). Augmented and virtual reality (AR/VR) for education and training in the AEC industry: A systematic review of research and applications. Buildings, 12, 1529.
\item Xu, J., \& Moreu, F. (2021). A review of augmented reality applications in civil infrastructure during the 4th industrial revolution. Frontiers in Built Environment, 7, 640732.
\item Arena, F., Collotta, M., Pau, G., \& Termine, F. (2022). An overview of augmented reality. Computers, 11(2), 28.
\item tom Dieck, M.C.; Jung, T.; Han, D.-I. Mapping requirements for the wearable smart glasses augmented reality museum application. J. Hosp. Tour. Technol. 2016, 7, 230–253.
\item https://www.apple.com/apple-vision-pro/specs/
\item Vlahakis, V., Ioannidis, M., Karigiannis, J., Tsotros, M., Gounaris, M.,
Stricker, D., Gleue, T., Daehne, P., \& Almeida, L. (2002).
Archeoguide: An augmented reality guide for archaeological sites.
IEEE Computer Graphics and Applications, 22, 52--60.
\url{https://doi.org/10.1109/MCG.2002.1028726}

\end{enumerate}
\end{document}